# Inverter Pulse Width Modulation Control Techniques for Electric Motor Drive Systems

**Prerit Pramod**, *Senior Member*, IEEE

Control Systems Engineering, MicroVision, Inc.
*Email*: preritpramod89@gmail.com; preritp@umich.edu; prerit_pramod@microvsion.com

**Abstract:** This paper provides a simple introduction to pulse width modulation control techniques used for the control of power converters in the context of electric motor drive systems. A summary of each technique is presented along with analytical models that provide intuitive insight and enable their rapid implementation for practical purposes.

## Introduction

The major hardware components typically employed in an electric motor drive system are shown in the figure below. Note that even though the figure illustrates a permanent magnet synchronous motor (PMSM) [1]–[5], the foregoing description of the inverter control techniques is applicable for other electric machine topologies including induction motors and switched reluctance motors [6]–[14], i.e., to any system that employs multi-phase voltage inverters.

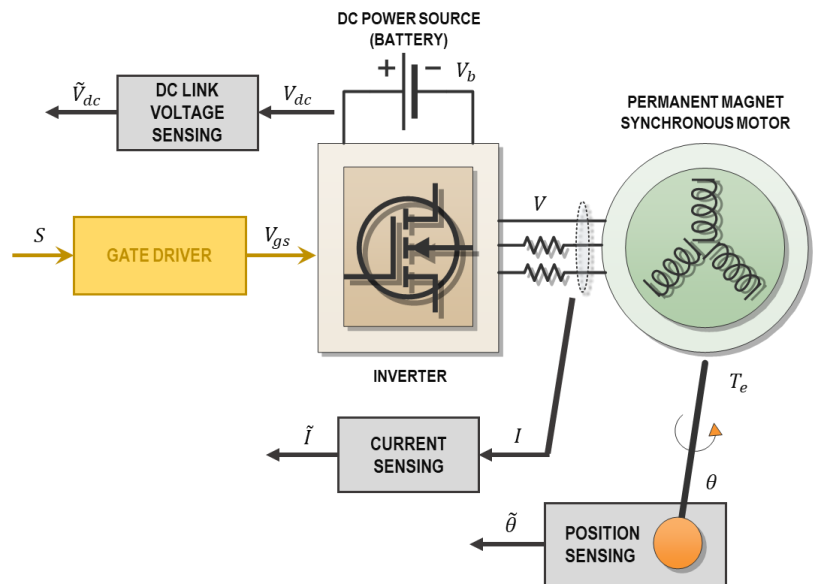

*Figure 1*: Permanent magnet synchronous motor drive system.

The power converter, which is an inverter in this case, is driven by a gate driver. The gate driver generates voltage signals to drive the gates of the individual switches within the inverter resulting in the generation of phase to ground voltages that are applied to the terminals of the machine. The inverter essentially converts the input DC voltage into voltage pulses through pulse width modulation (PWM) such that the average voltage during a given switching period equals the desired voltage command [15]. The motor then generates current and torque which are applied to the mechanical system within which the electric motor drive is used. The DC power source, which may be a battery, provides input power. The inverter input voltage at the bulk capacitor differs from the battery voltage due to the power input filtering circuit that precedes it [16]–[18]. Sensors for motor currents [19]–[26] and position [27]–[31] along with a DC link voltage measurement circuit are used to provide feedback for controlling the PMSM drive.

## Power Converter Control System

The input to the gate driver, which is essentially an amplifier, consists of switching instants specifying the times at which the different switches need to be turned on and off within a PWM period [32]–[37]. The gate drives utilize a timer unit along with the switching instants to generate and apply gate-to-source voltages to the gates of the individual switches which in-turn causes the switches to turn on and off. The overall block diagram of the power converter control system is shown in the figure below.

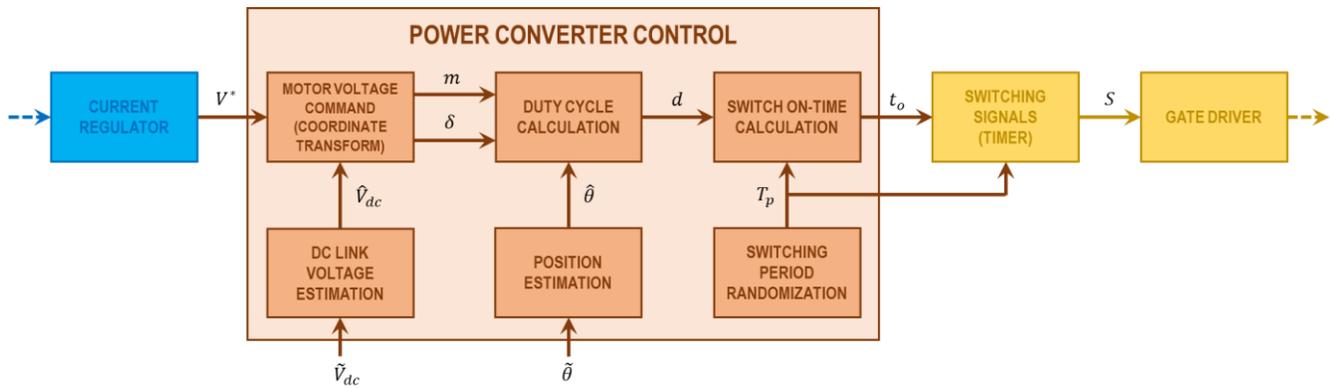

*Figure 2*: Power converter control system in electric motor drives.

As illustrated in the figure, the dynamic current controller which may be a feedback regulator [38]–[40] that acts on the error between the commanded and measured currents or a feedforward compensator [41]–[44] that employs an inverse model of the motor applied to the reference currents, sends synchronous frame voltage commands to the power converter control block. Note that the current commands are typically generated through capability limiting and power management algorithms for the motor drive system [45]–[50]. The current controller outputs voltage commands $V_{dq}^*$ which are converted to polar coordinates, i.e., into modulation index and phase advance, as (1).

$$m = \frac{\sqrt{V_d^{*2} + V_q^{*2}}}{\hat{V}_{DC}}$$
$$\delta = \tan^{-1}\left(\frac{V_d^*}{V_q^*}\right)$$

(1)

where $m$ is modulation index, $\delta$ is phase advance and $\hat{V}_{DC}$ is the estimated DC link voltage. Note that the DC link essentially refers to the voltage across the bulk capacitor which is located in close proximity to the power converter and acts as the primary power source for the drive system.

These signals are then used along with estimated electrical position to compute the duty cycles for the individual phase legs of the power converter. Different techniques for computing duty cycle are referred to as commutation techniques, which are the primary focus of this paper. Duty cycle calculation may be performed by using one of many pulse width modulation (PWM) techniques, which are described in detail in a later section.

While position estimation seems to be trivial on the face of it, in reality there are subtle aspects that are not understood by most engineers. I won't get into all the details here, but I will mention one aspect, namely the distinction between line and phase position signals. In actual software implementation and the end-of-line (EOL) calibration process, the position we estimated is referred to (at least by me) as line position. However, most of the time the stationary to synchronous (and inverse) reference frame transformations are specified in terms of phase position. This is not very obvious because the computations related to this is buried deep within specific components within the software. The line and phase electrical position signals are related as (2) for a three-phase machine.

$$\theta^l = \theta^p + v * \theta_0$$
$$\theta_0 = \frac{\pi}{6} \tag{2}$$

where the superscripts $l$ and $p$ represent line and phase respectively, $v$ is the electromechanical polarity of the machine (equal to +1 or -1 for positive or negative polarity respectively) and $\theta_0$ is a constant position offset. In this paper, everything is expressed in terms of the phase position signal.

The duty cycle signals which have values between zero and unity are converted into on and off times for the individual switches of the converter by multiplying them by the PWM period as (3).

$$t_{or}^h = T_p d_r$$
$$t_{fr}^h = T_p - t_{oa}^h \tag{3}$$

where $t_{or}^h$ and $t_{fr}^h$ represent the on and off times respectively of the upper switch in $r^{th}$ phase leg, $T_p$ is the PWM period and $d_r$ is the duty cycle. The PWM period is usually not a fixed number, rather a randomized signal around a nominal period. This randomization is referred to as dithering and is implemented for the purposes of spreading the noise frequency spectrum to avoid generating a pure tone at the switching frequency.

## Commutation Techniques
### *Sinusoidal Pulse Width Modulation (SPWM)*
The duty cycle for the $r^{th}$ phase of the machine for SPWM is expressed as (4).

$$d_r^S = \frac{1}{2}\left(1 + m * \sin\left(\hat{\theta} + \delta - (r-1)\beta\right)\right) \tag{4}$$

where $\hat{\theta}$ is the estimated electrical position and $\beta$ is the phase difference between adjacent phases.

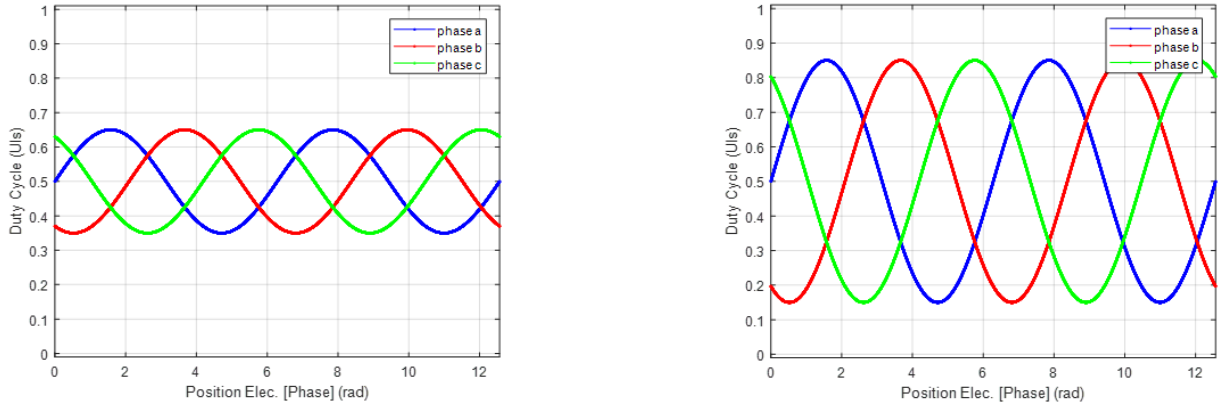

*Figure 3*: Sinusoidal PWM at (a) low (b) high modulation indices.

This is the most basic PWM technique and is rarely ever used for practical applications. The main reason behind this is that it produces sub-optimal DC bus voltage utilization. More specifically, if the SPWM technique is implemented, while the phase to ground voltages vary from zero to the DC link voltage, the resultant line to line voltages are limited to $\frac{1}{\sqrt{3}}$ of the maximum available voltage as shown in the figure below.

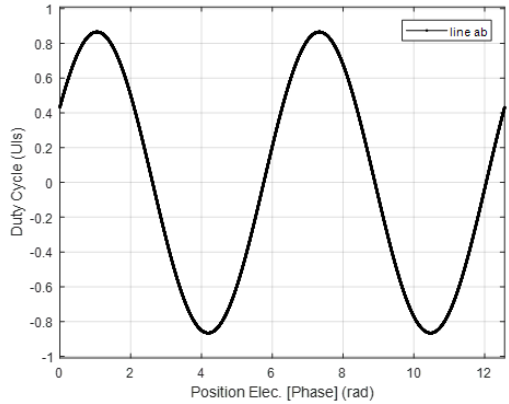

*Figure 4*: Line-to-line voltage for sinusoidal PWM technique.

In order to increase the line-to-line voltages, several different types of commutation techniques which modifications of the basic SPWM technique are essentially, may be implemented.

### Third Harmonic Injection Pulse Width Modulation (THPWM)
This technique involves adding a third harmonic component to the basic SPWM signals in order to maximize the bus voltage utilization. The duty cycle waveforms for THPWM for a three-phase machine are expressed as (5).

$$d_r^T = \frac{1}{2}\left(1 + m * \frac{2}{\sqrt{3}} * \left(\sin\left(\hat{\theta} + \delta - (r-1)\beta\right) + \frac{1}{6}\sin\left(3\left(\hat{\theta} + \delta - (r-1)\beta\right)\right)\right)\right) \tag{5}$$

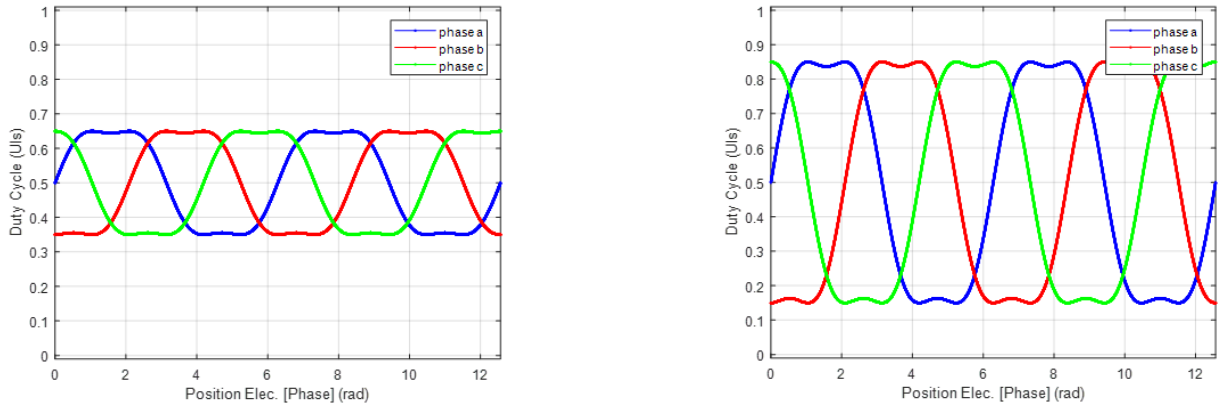

*Figure 5*: Third harmonic injection PWM at (a) low (b) high modulation indices.

The rationale behind adding a third harmonic is that these voltages get cancelled in the line-to-line for a three-phase machine, so the currents are ultimately unaffected. Due to the injection of this harmonic, the peak value of the fundamental component is now increased such that the DC bus voltage utilization is increased by a factor of $\frac{1}{\sqrt{3}}$.

### Discontinuous Space Vector Pulse Width Modulation (DPWM)

This technique basically "grounds" one of the phases through a modification of the SPWM technique and is therefore also called phase grounding. The duty cycles for DPWM are expressed as (6).

$$d_r^D = \frac{2}{\sqrt{3}}\left(\frac{1}{2}\left(1 + m * \sin(\hat{\theta} + \delta - (r-1)\beta)\right) - \min(d_a^S, d_b^S, \ldots d_q^S)\right) \tag{6}$$

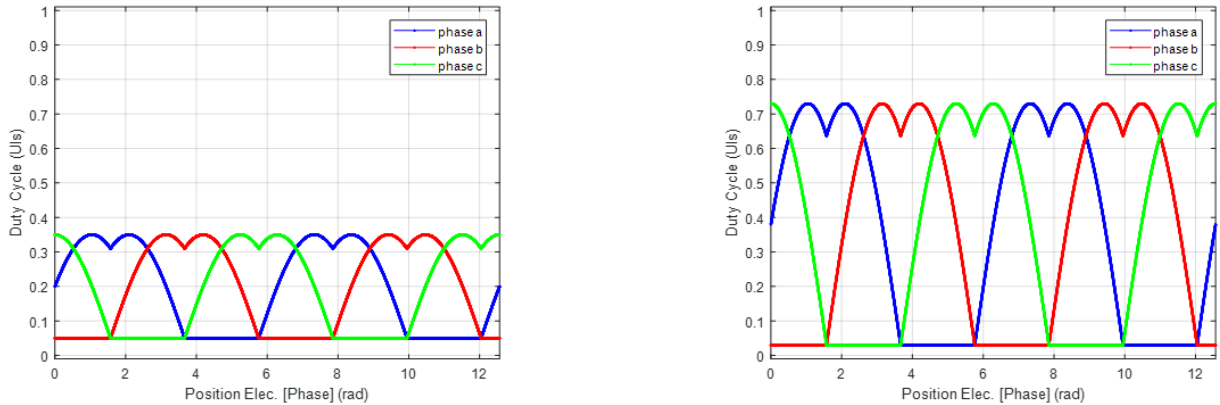

*Figure 6*: Discontinuous space vector PWM at (a) low (b) high modulation indices.

Although conceptually this is good for reducing switching losses, the non-linearity exhibited by switches at very low duty cycles causes issues and is eliminated by introducing an offset (known as commutation offset) that varies with modulation index in each of the phases as given in (7).

$$f(m) = \begin{cases} 1, & m < m_l \\ -\dfrac{m - m_h}{m_h - m_l}, & m_l \leq m < m_h \\ 0, & m \geq m_h \end{cases} \tag{7}$$

A graphical representation of the commutation offset as a function of modulation index is shown in the figure below.

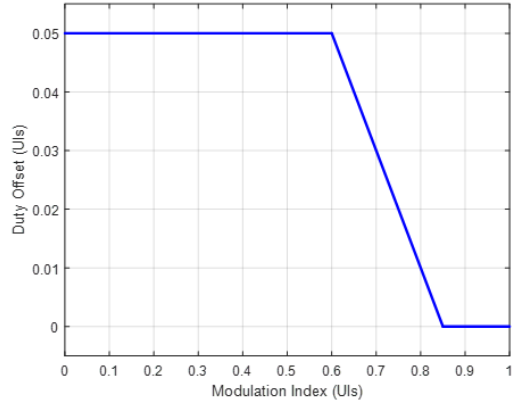

*Figure 7*: Commutation offset variation with modulation index.

The duty cycle calculation for the resulting DPWM technique with offset may be expressed as (8). Due to the commutation offset, this technique is also referred to as phase offset grounding.

$$d_r^p = \frac{2}{\sqrt{3}} \left( \frac{1}{2} \left( 1 + m * \sin\left(\hat{\theta} + \delta - (r-1)\beta\right)\right) - \min\left(d_a^S, d_b^S, \dots d_q^S\right)\right) + d_0 f(m) \tag{8}$$

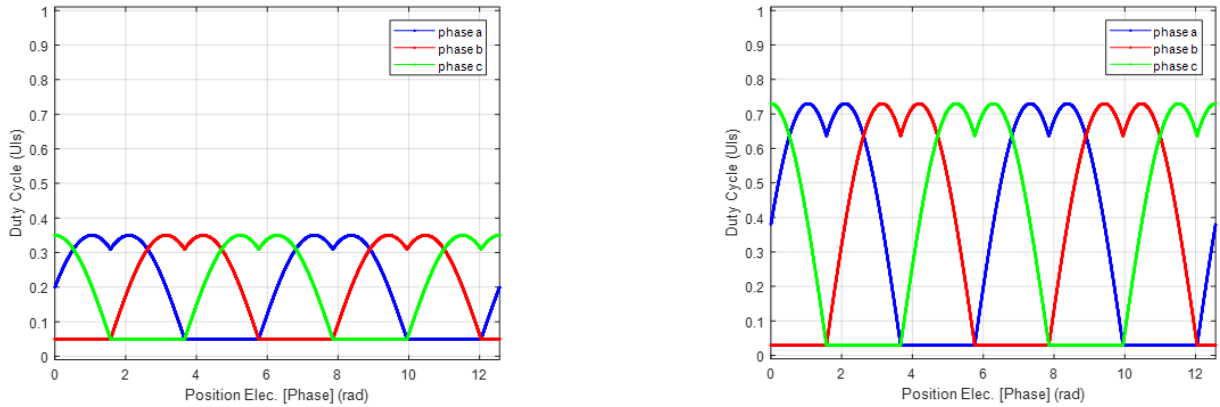

*Figure 8*: Discontinuous space vector PWM with offset at (a) low (b) high modulation indices.

As in the case of THPWM, the DPWM techniques also provide higher bus voltage utilization which can be observed from the corresponding line-to-line voltages shown in the figure below.

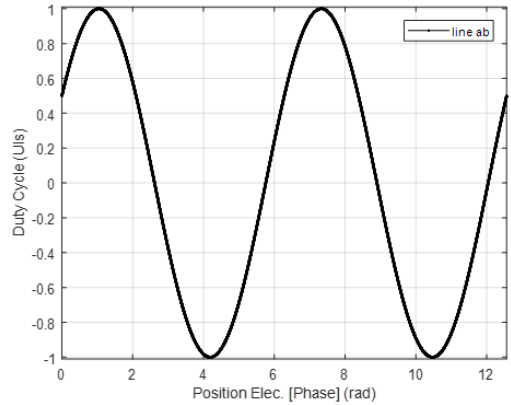

*Figure 9*: Line-to-line voltage for discontinuous space vector PWM technique.

As mentioned earlier, the software implementation of the different commutation techniques is in terms of the line position signal even though the expressions presented here are in terms of phase position. The figure below illustrates the difference between the implementation of the duty cycle calculation in terms of these two position signals for a positive electromechanical polarity machine.

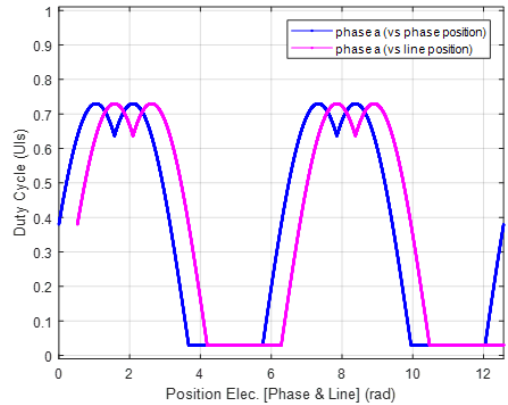

*Figure 10*: Discontinuous space vector PWM duty cycle variation with phase and line position.

### *Continuous Space Vector Pulse Width Modulation (CPWM)*

This is the most commonly used technique for commutation of synchronous motor drives because of its symmetric nature and maximum bus voltage utilization. The duty cycle calculation for CPWM is given in (9).

$$d_r^C = d_r^D + \frac{1}{2}\big(1 - \max\big(d_a^D, d_b^D, \dots d_q^D\big)\big) \tag{9}$$

The CPWM technique is often called Space Vector PWM (SVPWM), although this is not strictly correct terminology.

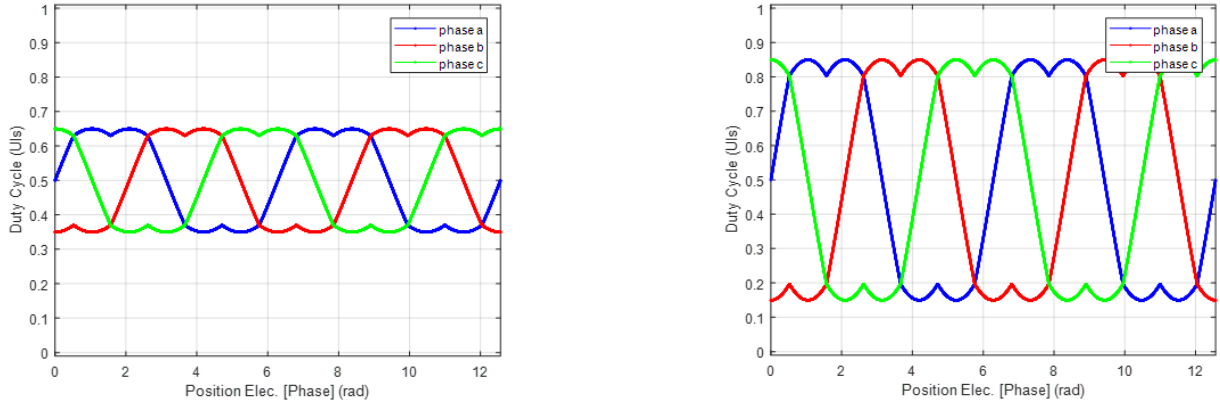

*Figure 11*: Continuous space vector PWM at (a) low (b) high modulation indices.

Due to the symmetric nature of the CPWM technique, i.e., due to its symmetry about 0.5 duty, the resultant line to line switching signals have a dominant frequency component at twice the switching frequency (equivalent to the PWM period) for a specific range of modulation indices. This is different from the DPWM technique where the dominant frequency content is at the switching frequency.

### Adaptive Pulse Width Modulation (APWM)

This technique is a hybrid of the DPWM and CPWM techniques and is obtained as a blend of the two as given in (10).

$$d_r^A = (1 - b(m))d_r^C + b(m)d_r^D \tag{10}$$

The blending factor is varied as a function of the modulation index as expressed in (11).

$$b(m) = \begin{cases} 1, & m < m_l \\ -\dfrac{m - m_h}{m_h - m_l}, & m_l \leq m < m_h \\ 0, & m \geq m_h \end{cases} \tag{11}$$

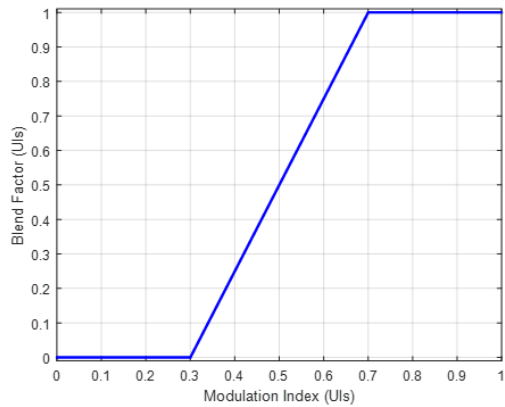

*Figure 12*: Scale factor for blending continuous and discontinuous PWM.

The main idea behind the APWM technique is to combine the benefits of both CPWM and DPWM, i.e., shift of noise frequency content to a higher range and lower switching losses at lower and higher modulation indices respectively. The duty cycle waveforms for the APWM technique are shown in the figure below.

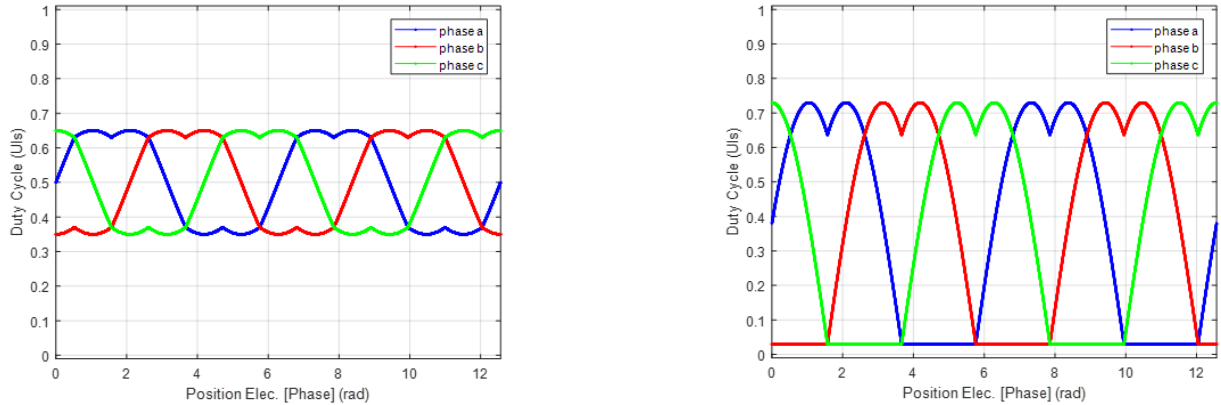

*Figure 13*: Adaptive space vector PWM at (a) low (b) high modulation indices.

## Conclusion

This paper provides a summary of different power converter commutation techniques most commonly used for the control of voltage source inverters. Although the description provided is in the context of electric motor drives, the concepts are not so limited and possess utility and applicability for a plethora of applications, most generally for multi-phase converters. The analytical closed form representations enable the rapid deployment of these methods for practical use.